\documentclass[pre,preprintnumbers,amsmath,amssymb,floatfix,12pt,authoryear]{revtex4}

\usepackage{graphics}
\usepackage{amsmath}
\usepackage{graphicx,color}
\usepackage[latin1]{inputenc}
\usepackage[T1]{fontenc}
\usepackage{ae,aecompl}

\renewcommand{\H}{\mathcal{H}}

\begin{document}

\title{Probing rare physical trajectories with Lyapunov weighted dynamics}

\author{Julien~Tailleur}
\email{tailleur@pmmh.espci.fr}

\author{Jorge~Kurchan}
\email{jorge@pmmh.espci.fr}

\break

\affiliation{ Laboratoire PMMH (UMR 7636 CNRS;ESPCI;P6;P7)\\10, Rue Vauquelin, 75231 Paris CEDEX 05, France}

\maketitle

{\bf The transition from order to chaos has been a major subject of
research since the work of Poincar\'e, as it is relevant in areas
ranging from the foundations of statistical physics to the stability
of the solar system\cite{LL,Ott93}. Along this transition,
atypical structures like the first
chaotic regions to appear, or the last regular islands to survive,
play a crucial role in many physical situations. For instance,
resonances and separatrices determine the fate of planetary
systems\cite{Laskar89,MuHo}, and localised objects like 
solitons and breathers provide mechanisms of energy transport in
nonlinear systems such as Bose-Einstein condensates and biological
molecules\cite{Tanos,CB}. Unfortunately, despite the fundamental
progress made in the last years, most of the numerical methods to
locate these 'rare' trajectories are confined to low-dimensional or
toy models, while the realms of statistical physics, chemical
reactions, or astronomy are still hard to reach. 
Here we implement an efficient method that allows one to
work in higher dimensions by selecting trajectories with unusual
chaoticity. As an example, we study the Fermi-Pasta-Ulam nonlinear
chain in equilibrium\cite{CB} and show that the algorithm rapidly
singles out the soliton solutions when searching for trajectories with
low level of chaoticity, and chaotic-breathers in the opposite
situation. We expect the scheme to have natural applications in
celestial mechanics and turbulence, where it can readily be combined
with existing numerical methods.}

Complex dynamical systems are generically chaotic: two nearby
trajectories  diverge
exponentially with time, at a rate given by the
 Lyapunov exponent: $|\delta x(t)|=|\delta x(0)| \exp(t
\lambda_{\text{orb.}})$. If chaoticity is sufficiently strong, almost every
initial condition leads in the large-time limit to the same exponent
$\lambda = \underset{t\to\infty}{\text{lim}}
\lambda_{\text{orb.}}(t)$. Such an averaged description is not
complete, as in many systems some trajectories will present atypically
chaotic or regular behaviour extending (in the future or in the past)
for very long periods -- even forever -- depending on their initial
conditions.

A case in point is that of  a planetary system
\cite{Laskar89,Murray99,MuHo}:  of the possible
sets of masses and orbital elements compatible with observational error,
only some correspond to a situation not involving a planet ejection
in the  recent past or near future.
Another example, to which we shall return below, is that of
nonlinear media, where spatially localised soliton and
`breather' solutions 
(which have atypical Lyapunov exponents) are reached only from very
specific initial conditions\cite{CB,Tanos}.

In order to describe quantitatively the distribution of Lyapunov
exponents of different trajectories, two closely related approaches,
both inspired in thermodynamics, have been proposed in the past. In
one approach, the sampling is over trajectories of fixed time and
different initial conditions~\cite{Grassberger}, while in the other
 the dynamics is perturbed with a small random noise, and the sampling
is made over different noise realisations~\cite{Benzi}. Lyapunov weighted
dynamics (LWD) (a method proposed in the context
of chemical reactions~\cite{STN,STN1}) can be modified to
 practically implement both these approaches 
-- although here for brevity we discuss only the latter (see `Method'). A
population of walkers evolves in phase space with Hamiltonian
dynamics, each one perturbed by an independent weak random force,
which may be energy-conserving or not. The walkers carry with them
a `Lyapunov' vector that itself evolves as the separation between two
nearby trajectories. After every time interval,
walkers are cloned (or killed) with a rate proportional to ($\alpha$
times) the stretching rate of their Lyapunov vector.  The end effect
of the cloning is that orbits are now weighted according to their
chaoticity.
 
A positive (negative)
value of $\alpha$ tends to favour orbits with large (small) Lyapunov
exponent. In fact, this procedure counts the trajectories with a
weight $\exp(\alpha \lambda_{\text{orb.}}
t)$ (See supplementary notes). Within the formalisms mentioned above,
 $\alpha$ and $\lambda$ are  conjugate
variables, just as inverse temperature and energy are in a
thermodynamic problem: fixing $\alpha$ we obtain a typical value of
$\lambda$, together with its probability and the trajectories that
contribute.

In what follows we shall apply our method to several examples of
increasing complexity.

\begin{minipage}{\columnwidth}
  {\bf \large Separatrices}\\
  Separatrices are the phase-space frontiers between regions with
  different dynamical behaviours.  They play an important role because
  they are the cradle of chaos.  In regular systems, walkers weighted
  with $\alpha=1$ can be shown to converge there, and to populate them
  uniformly. This is illustrated in figure 1 for the simple example of
  a double well with potential energy $V(q)=p^2/2+q^4-2 q^2$.
  Particles can either librate within one of the wells, or oscillate
  between them. The separatrix is the curve emanating from the saddle
  point that limits these two regimes. Applying the dynamics with
  $\alpha=1$, the clones slowly diffuse in energy because of the noise
  until they converge to the separatrix, where they multiply more
  favourably. Clones that subsequently diffuse away die without
  leaving offspring.
\end{minipage}

\begin{figure}[h]
  \begin{center}
    \includegraphics[width=.5\columnwidth]{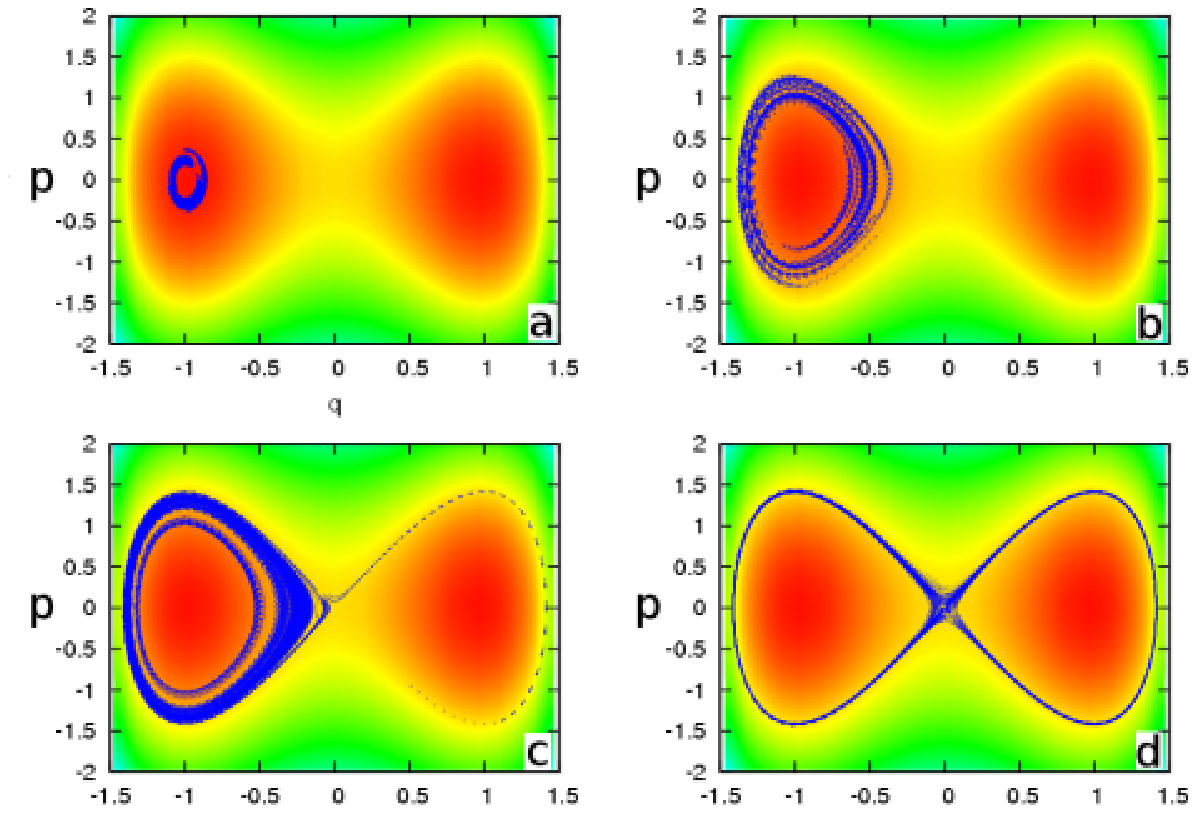}
    \caption{{\bf Convergence towards the separatrix.}\\ The color
      code represents the value of the energy. Starting in one well
      ({\bf a}, $t \sim 3000$), the walkers first diffuse in energy
      ({\bf b}, $ t \sim 11450$) until they reach the separatrix ({\bf
        c}, $t\sim11725$), where they settle ({\bf d}, $t>12000$).
      Two thousands walkers were run with $\epsilon=10^{-5}$ and
      $\alpha=1$.  }
  \end{center}
\end{figure}

Let us now consider what happens as chaos sets in.
 We illustrate this with the classical example of the
`Standard Map':
\begin{equation}
\label{eqn:standard}
p_{n+1}=p_n - \frac{k \delta}{2 \pi} \sin(2 \pi q)\qquad q_{n+1} = q_n
+ \delta p_{n+1}
\end{equation}
It represents the evolution of a free rotor regularly kicked with a
constant force of fixed orientation \cite{Ott93}. In the limit $\delta
\to 0$, it reduces to the evolution of a pendulum, and is thus
integrable. For $\delta>0$, it becomes chaotic, the more so the larger
 $\delta$ and $k$.  Applying the
algorithm for $\alpha=1$ close to integrability (figure
2.a), the walkers concentrate on the
unstable manifold emanating from the fixed point, revealing the
typical features of the homoclinic tangle.  When chaos is increased
--- for larger $\delta$ --- many secondary structures become apparent
(figure 2.b). Around the main resonant
island, a stochastic layer is now the most chaotic structure of the
phase space, and will consequently be the favoured target of the
walkers. Nevertheless,  starting from a given configuration,
 the walkers first converge to the closest stochastic layer,
where they stay for some time.  In other words, chaotic layers are the
metastable states of the $\alpha=1$ dynamics.

\begin{figure}
  \begin{center}
    \includegraphics[width=1\columnwidth]{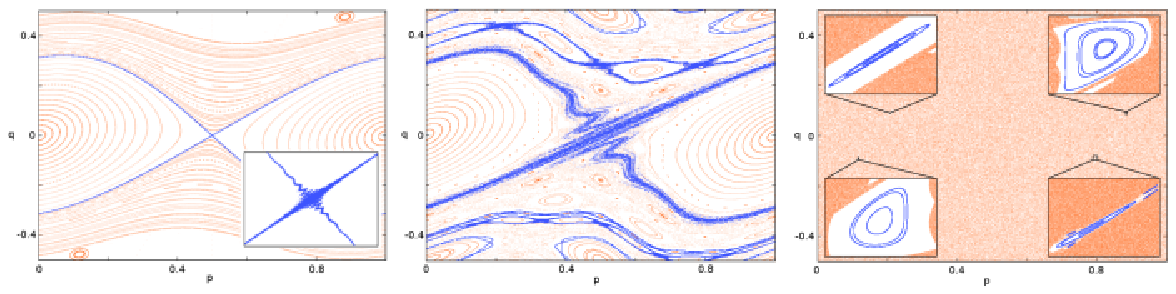}
    \caption{{\bf The Standard Map}\\
      The standard map trajectories are represented in orange. In blue,
      a thousand walkers at time $T=10,000$ evolving with
      $\epsilon=10^{-16}$. {\bf a)} Quasi-integrable case:
      $\delta=0.41$, $k=1$. The map is very slightly chaotic, and the
      walkers converge for $\alpha=1$ to the homoclinic orbit emanating
      from the unstable fixed point. The inset is enlarged seventy-five
      times. {\bf b} Secondary structures: $\delta=1$, $k=1$. Several
      separatrices are revealed, starting from different regular islands
      with $\alpha=1$. {\bf c} The last four regular islands in a
      strongly chaotic case~\cite{Loco}: $\delta=1$, $k=7.7$
      $\alpha=-1$. The insets are zoomed between 25 times (bottom right)
      and 150 times (top) and are centered around: $(0.207;0.09)$,
      $(0.883;0.09)$, $(0.116;-0.09)$ and $(0.8;-.09)$.}
  \end{center}
\end{figure}

For a strongly chaotic system, we may wish to explore hidden
regular structures~\cite{Loco} lost in the chaotic sea. Choosing
 $\alpha=-1$, we bias the measure in favour of regular orbits,
thus revealing the --- otherwise invisible --- last resonant
islands (see figure 2.c).

\begin{figure}
  \includegraphics[width=.6\columnwidth]{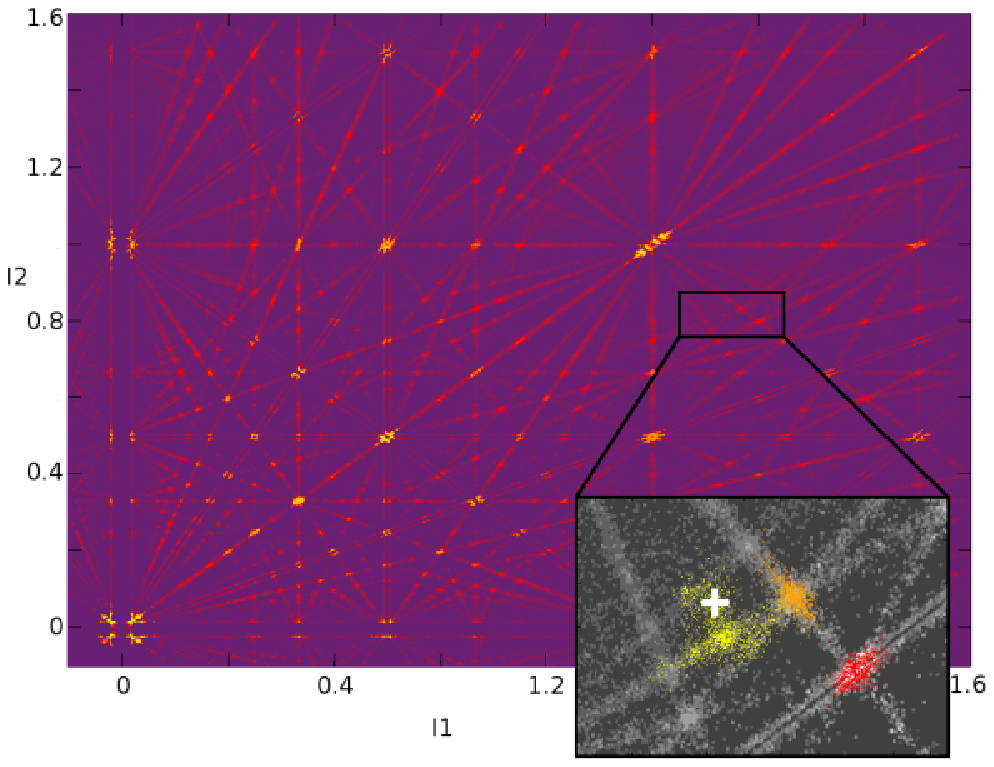}
  \caption{{\bf Arnold web and convergence.} \\
    Arnold Web for the Hamiltonian (2) with $\mu=10^{-3}$, very close to
    integrability. Lighter regions have stronger chaoticity.  In the
    inset: the evolution of a set of clones in three successive times,
    starting from the point indicated by the white cross. Yellow for
    $0<t<160,000$, orange for $160,000<t<2,000,000$ and red for
    $2,100,000<t<2,200,000$.  The large picture was composed by
    repeating the procedure shown in the inset for several initial
    conditions. Note that although the evolution of the clones
    resembles Arnold diffusion, it is in fact noise-driven, and hence
    orders of magnitude faster.}
\end{figure}

{\bf \large Arnold Diffusion}

An integrable Hamiltonian system has a constant of motion for every
degree of freedom. The phase space is filled with
invariant tori on which the motion is quasi-periodic, with frequencies
$\omega_1\cdots \omega_N$. Under a small perturbation,
phase space is qualitatively unchanged,  except for a chaotic
region in the neighbourhood of `resonances', defined as
  the points satisfying the
Diophantine relation: $\sum n_i \omega_i=0$ with $n_i$ integer. In
systems with more than two degrees of freedom, this `chaotic web'
leads to global diffusion in phase-space \cite{Ott93,LL}.
  This phenomenon, predicted
by Arnol'd in 1964 \cite{Arnold64}, is called Arnol'd diffusion and
the stochastic web is consequently called Arnol'd web.
Localising such  diffusive part of phase space is of great
interest for many systems, for instance in synchrotron experiments
\cite{Laskar00}, plasma physics or celestial mechanics.
 Laskar~\cite{Laskar93} developed a successful method, based on
 Fourier analysis, to map completely the  regular part of phase-space.
Here we are interested in going straight to the chaotic regions, which
 may be very thin and hard to find.
Applying LWD with $\alpha > 0$,
 the walkers  first diffuse in phase space thanks to 
the stochastic noise, until
they reach chaotic structures, where they multiply and settle. As a
benchmark, we constructed the Arnol'd web for the 
Hamiltonian~\cite{Froeschle00}:

  \begin{equation}
    \H=\sum_{i=1}^N \frac{p_i^2}{2} + p_{N+1} + \frac{\mu}{\sum_{i=1}^N
      \cos q_i + \cos t + N +2 }
    \label{H}
  \end{equation}

Consider first the two-dimensional case $N=2$. When $\mu=0$, this
Hamiltonian is integrable, and the resonance lines are given by $n_1
p_1+ n_2 p_2 = n_3$. For very small $\mu$ the chaotic set, shown in
figure 3, is concentrated around the resonance lines.  On the inset,
one can see the path followed by the walkers that started in a random
initial condition: they diffuse and settle in successive chaotic
regions. The important point is that although the trajectory of the
clones mimics Arnold diffusion, it happens in a timescale orders of
magnitude shorter, as it is driven by noise. On the inset of figure 4
we show that each time the clones find a resonance there is an
inflection in the finite-time Lyapunov exponent.  Figures like 3 have
been obtained in \cite{Froeschle00} by computing the finite-time
Lyapunov exponents of trajectories starting in points on a grid in
$(p_1,p_2)$.  Here, instead, we run LWD as in the inset, starting from
several initial conditions.

\begin{figure}
  \includegraphics[width=.5\columnwidth]{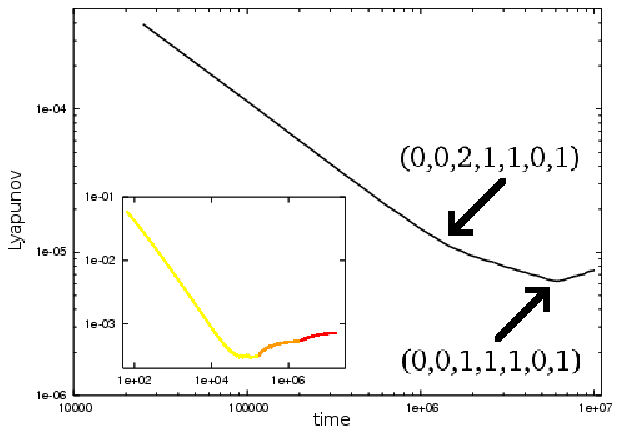}
  \caption{ {\bf Finding resonances in higher dimensions. }\\ Main figure: a $6$
    degree of freedom system with Hamiltonian (2) and $\mu=5.10^{-6}$,
    starting from a random initial condition.  The two inflection points
    indicated by the arrows correspond to times at which the clones
    reached the resonances labeled $(n_1,n_2,n_3,n_4,n_5,n_6,n_7)$. In
    the inset: the same plot for the dynamics in the inset of figure
    3; yellow, orange and red correspond to the
    same three times shown there.}
\end{figure}

In higher dimensions, the Arnold web is both difficult to represent
and impossible to map exhaustively using a grid. There is no problem,
however, in applying LWD as described above. To illustrate this, in
figure 4 we show for the $6$ degree of freedom
version of (\ref{H}) the running-time evaluation of the clones'
Lyapunov exponent starting from a random configuration. Using a
separate programme, we have checked that the inflections indeed
occured when the clones found the resonances indicated with
arrows. Few-planet systems are clearly within reach, and LWD could be
used to locate nearby chaotic trajectories in agreement with
experimental data, in the spirit of \cite{LaskarMercury}.

\begin{figure}
  \includegraphics[width=\columnwidth]{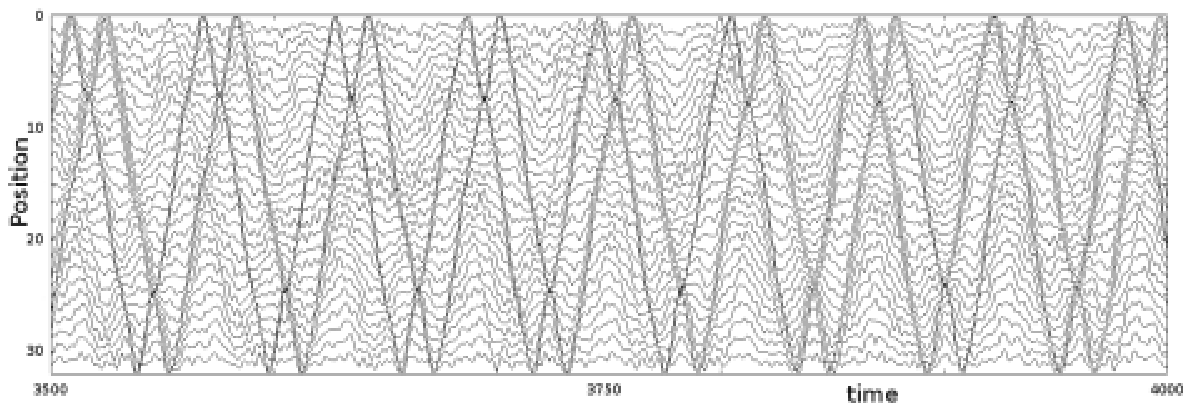}
  \caption{{\bf Finding solitons} \\
    Energy-conserving ($E=1$) dynamics of the Fermi-Pasta-Ulam chain
    with fixed ends, and strongly negative $\alpha$, obtained starting
    from microcanonical equilibrium.  The plot shows the position of
    the particles versus time.  Several kinks bouncing from an end to
    the other of the chain are clearly visible. The Lyapunov exponent
    is half the typical value. (For clarity, the particles' average
    position has been arbitrarily displaced).}
\end{figure}

{\bf \large Breathers and Solitons}

As an example with many degrees of freedom, consider the
 Fermi-Pasta-Ulam chain, defined by the Hamiltonian:
\begin{equation}
H=\sum_i \left( \frac{1}{2} p_i^2+ 
\frac{1}{2} (x_i-x_{i+1})^2 + \frac{1}{40} (x_i-x_{i+1})^4  \right)
\end{equation}
This chain is known to have localised soliton excitations, as well
as the so-called `chaotic breathers'~\cite{CB,Tanos}.
These excitations are unstable, but can be observed for some time
starting from suitable initial conditions.
Solitons configurations have atypically low chaoticity, while, perhaps
more surprisingly, chaotic breathers are indeed very chaotic~\cite{CB}.

We have run an energy-conserving version of LWD on the chain starting
from a microcanonical equilibrium configuration. The results, for an
energy density $E=1$, corresponding to the `transition
region'~\cite{CB}, are shown in figures 5 and 6.  When biasing to
obtain a Lyapunov exponent three times larger than typical, we observe
a chaotic-breather automatically emerging, thus showing that the large
Lyapunov configurations are dominated by them.  Conversely,
constraining the Lyapunov exponent to be a half of the typical value
results in a few-kink configuration, as is clearly visible in figure
5.

\begin{figure}
  \includegraphics[width=.7\columnwidth]{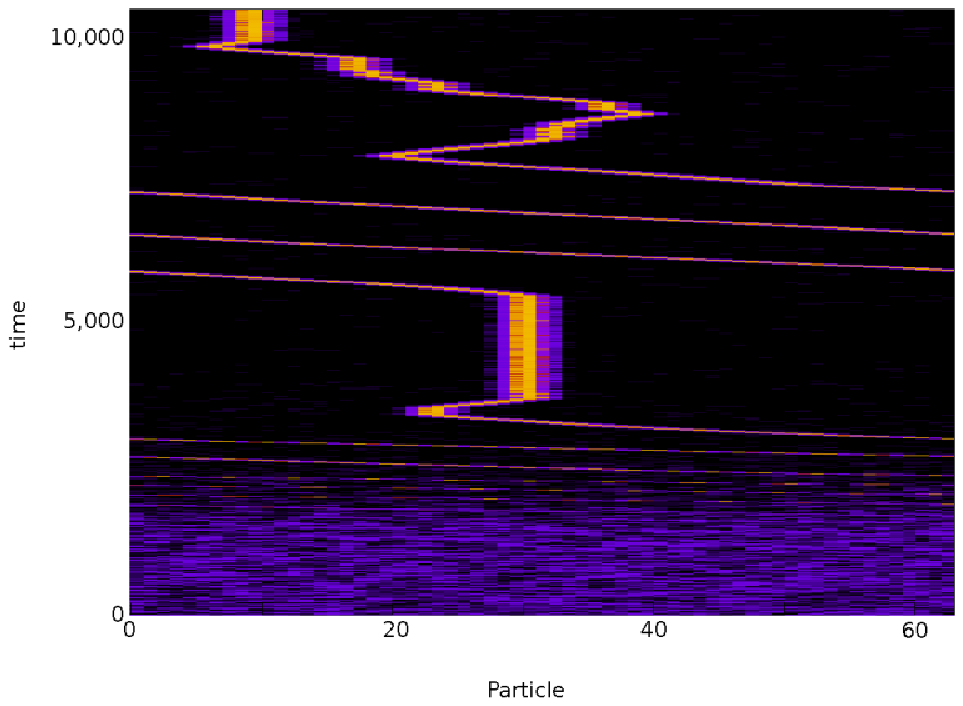}
  \caption{{\bf Finding a chaotic breathers.}\\
    Energy-conserving (energy density $E=1$) dynamics of the
    Fermi-Pasta-Ulam chain with periodic boundary conditions and
    large, positive $\alpha$, starting from microcanonical
    equilibrium.  The configuration evolves towards a single chaotic
    breather, whose Lyapunov is three times the one of a typical
    equilibrium trajectory. The colour code corresponds to the energy
    of the site.  }
\end{figure}

{\bf \large Conclusion}

Let us conclude by suggesting that the special trajectories described here
play the same  role as the reaction paths in physical chemistry, whose
efficient detemination is currently a field of intense activity~\cite{tps}. 
 In  all cases one 
has to deal with  events that are important but happen rarely, 
 so  they are best studied with a  dynamics that is biased to
enhance the probability of their occurrence  --
although of course one is ultimately interested in the properties of
the true, physical dynamics.

{\bf \large Method}
A population of walkers with phase-space positions
${\bf x^a} \equiv ({\bf q^a},{\bf p^a})$ 
is evolved with the standard Hamiltonian dynamics $\dot {\bf q^a}
= {\bf p^a}$, $\dot {\bf p^a} = -\nabla V({\bf q^a
})+\sqrt{\epsilon} {\bf \eta}(t)$, where $\eta$ is a white noise of unit
variance.
Every clone has a `companion'
 which starts a small distance $ \delta  {\bf x^a(0)}$
away, and evolves with the same noise. After a time interval $ \delta t$,
when  the positions
of clone and companion are 
$[({\bf x}^a ( \delta t),{\bf x }^a 
( \delta t)+\delta {\bf x}^a ( \delta t)]$, and the separation ratio
 $p_a= \left(\frac{||\delta {\bf x}^a 
( \delta t)||}{||\delta {\bf x}^a (0)||}\right)$
  {\em i)}  their mutual separation   is normalised
 $\delta {\bf x}^a \rightarrow \delta {\bf x}^a/p_a $,
 {\em ii)}  
if 
$(p_a)^\alpha>1$ the  clone $a$ is copied (i.e. replaced by two clones)
with probability $(p_a)^\alpha-1$, while if
$(p_a)^\alpha<1$ it is killed  with probability $1-(p_a)^\alpha$.
Each clone-companion pair subsquently evolves with a different noise 
realisation. 
 An overall
cloning rate is then applied to keep the population constant.
In the energy-conserving variant, which we applied to the particle chain,
 the only difference is that the noise vector is tangential to
the energy surface.

\begin{acknowledgments}
We would like to thank Giovanni Galavotti,
 Jacques Laskar,  Stefano Ruffo and Sorin Tanase-Nicola 
for very  useful discussions. 
\end{acknowledgments}

\end{document}

\section{Supplementary Notes. Cloning rate and probability distribution function}

\subsection{Evolution of walkers}

Hamilton's equations of motion with noise: $ {\bf q} = {\bf p}$,
$\dot {\bf p} = -\nabla V({\bf q})+\sqrt{\epsilon} {\bf
\eta}(t)$ can be written in a compact form:
\begin{equation}
\begin{aligned}
\label{eqn:Ham}
\dot x_i &= -\omega_{ij} \ddp{\H}{x_j}+D_{i} \eta_i \\
\end{aligned}
\end{equation}
where ${\bf x} = (q_1 \cdots q_N,p_1 \cdots p_N)$,
$D_i=(0\cdots0,1\cdots 1)$, $\omega={\tiny \begin{pmatrix} 0&
-1\\1&0\end{pmatrix} }$ and  repeted indices are summed over. The
divergence of nearby trajectories is measured by the Lyapunov vector
${\bf u}$ which evolves as
\begin{equation}
\label{eqn:HamLin}
\dot u_i = -A_{ij} u_j \qquad\text{where}\qquad A_{ij}=
\omega_{ik} \ddpp{\H}{x_k}{x_j}
\end{equation}
In terms of the unit vector ${\bf v}=\frac{{\bf u}}{|{\bf u}|}$,
the norm  of ${\bf u}$ is given by 
\begin{equation}
|{\bf u}(t)|=|{\bf u}(0)| e^{-\int_0^t v_i A_{ij} v_j \dt}
\end{equation}
while ${\bf v}$ evolves as
\begin{equation}
\label{eqn:HamLin2}
\dot v_i = -A_{ij} v_j + v_k A_{kl} v_l v_i
\end{equation}

The generating function of the largest Lyapunov exponent at time $t$
consequently reads
\begin{equation}
\langle e^{\alpha \lambda t} \rangle_{\text{traj.}} = 
\langle e^{-\alpha\int_0^t v_i A_{ij} v_j \dt } \rangle_{\text{traj.}}
\end{equation}
where the average is over trajectories given by (\ref{eqn:Ham}) and
(\ref{eqn:HamLin2}), and over the noise realisations.
Using standard methods, this average can be written
\begin{equation}
\langle e^{\alpha \lambda t} \rangle_{\text{traj.}} = \int \text{d}{\bf x}_t 
\text{d}{\bf v}_t \;\;\; e^{- t \big(H_{FP}
+\alpha v_i A_{ij} v_j\big) } P_0(x_0,v_0) 
\end{equation}
where $P_0$ is the initial probability distribution and $({\bf
x}_t,{\bf v}_t)$ is the end point of the trajectories. The operator 
$H_{FP}$ reads
\begin{equation}
\label{eqn:HK}
H_{FP}= -\frac{\epsilon}{2} \frac{\partial^2}{\partial p_i^2
} +p_i \ddp{}{q_i}-\ddp{V(q)}{q_i}\ddp{}{p_i} -\ddp{}{v_i}\left(A_{ij}v_j - 
v_k A_{kl} v_l v_i \right)
\end{equation}
Equation (\ref{eqn:HK}) corresponds to an evolution:
\begin{equation}
\ddp{P}{t}=-(H_{FP}+\alpha v_i A_{ij} v_j) P
\end{equation}
It does not conserve probability:
\begin{equation}
\int \text{d}{\bf x} \text{d} 
{\bf v} \ddp{P}{t}=-\alpha \int
 \text{d}{\bf x} \text{d} {\bf v} \;\;  v_i A_{ij} v_j P
\end{equation}
This simply says that if one represents $P(t)$ by a
distribution of walkers, each one is cloned at every time step
with a rate $-\alpha A_{ij} v_i v_j$, nothing but $\alpha$ times the
stretching rate of ${| \bf u(t)|}$. 

Clearly this derivation is not restricted to Hamiltonians of the form
$p^2/2+V(q)$ and is valid also for the time dependent case.
Furthermore, maps like the standard map can always be written as a
continuous time dependent Hamiltonian evolution:
\begin{equation}
\begin{aligned}
\dot q &=0 \qquad \dot p = \frac{k \delta}{2 \pi} 
\sin (2\pi q)\qquad \text{for even time intervals} \\ 
\dot q &=p \qquad \dot p = 0\qquad \qquad \qquad
\text{for odd time intervals} \\ 
\end{aligned}
\end{equation}
and the derivation above applies to them.

\subsection{ Energy conserving noise} For a Hamiltonian of the form
$\frac{{\bf p}^2}{2}+V({\bf q})$, we implement an energy conserving
stochastic dynamics by using a noise vector tangential to the energy
surface:
\begin{equation}
\eta_i=\sum_j \left(\delta_{ij} - \frac{p_i p_j}{{\bf p}^2}\right) \rho_j
\end{equation}
where $\rho_j$ are independent white noises. When this equation is
understood in the Stratonovich sense, it leads to microcanonical
measure.

\subsection{ Legendre transform }

Let us show at that time that by fixing $\alpha$ one can select -- in
the long time limit -- trajectories with a desired Lyapunov
exponent. By definition the generating function of the Lyapunov
exponent is given by
\begin{equation}
\langle e^{\alpha \lambda t} \rangle_{\text{traj.}} = \int \text{d} \lambda \; 
\rho(\lambda,t) e^{\alpha \lambda t}
\end{equation}
where $\rho(\lambda,t)$ is the probability density of $\lambda$ at time
$t$. As is well-known, $\rho(\lambda)$  can be written for large times as a 
large deviation function:
$\rho(\lambda,t) \simeq e^{t f(\lambda)}$. 
By saddle point evaluation one then gets
\begin{equation}
\langle e^{\alpha \lambda t} \rangle_{\text{traj.}}
 \simeq e^{t(\alpha \lambda^* +f(\lambda^*))}
\end{equation}
where $\lambda^*$ satisfies $f'(\lambda^*)=-\alpha$. The dominating
orbits indeed have $\lambda=\lambda^*$, and $(\alpha \lambda^*
+f(\lambda^*))$ is the typical cloning rate.
Thus, from $\lambda^*$ and the cloning rate, which are given by the algorithm,
one can reconstruct $\rho(\lambda)$.